# Atmospheric Oxygen Binding and Hole Doping in Deformed Graphene on a SiO$_2$ Substrate

Sunmin Ryu,[1,†] Li Liu,[2,†] Stephane Berciaud,[2,‡] Young-Jun Yu,[3] Haitao Liu,[2] Philip Kim,[3] George W. Flynn[2]* and Louis E. Brus[2]*

[1]Department of Applied Chemistry, Kyung Hee University, Yongin, Gyeonggi 446-701, Korea
[2]Department of Chemistry, Columbia University, New York, NY 10027, USA; ‡present affiliation - IPCMS (UMR 7504), Université de Strasbourg and CNRS, F-67034 Strasbourg, France
[3]Department of Physics, Columbia University, New York, NY 10027, USA
*E-mail: gwf1@columbia.edu; leb26@columbia.edu
†These authors contributed equally to this study

**Abstract**

Using micro-Raman spectroscopy and scanning tunneling microscopy, we study the relationship between structural distortion and electrical hole doping of graphene on a silicon dioxide substrate. The observed upshift of the Raman G band represents charge doping and not compressive strain. Two independent factors control the doping: (1) the degree of graphene coupling to the substrate, and (2) exposure to oxygen and moisture. Thermal annealing induces a pronounced structural distortion due to close coupling to SiO$_2$ and activates the ability of diatomic oxygen to accept charge from graphene. Gas flow experiments show that dry oxygen reversibly dopes graphene; doping becomes stronger and more irreversible in the presence of moisture and over long periods of time. We propose that oxygen molecular anions are stabilized by water solvation and electrostatic binding to the silicon dioxide surface.

## Introduction

Graphene, a single sheet of graphite, has been the subject of intensive research owing to its potential application in electrical devices,[1] flexible and transparent electrodes,[2] ultrathin membranes,[3] and various nanocomposites.[4] Recent reports of efficient chemical growth[5,6] and band-gap tuning in double-layered graphene[7] have expanded our ability to synthesize and control graphene for these applications.[8] Initial reports of thickness-dependent chemical reactivity,[9,10] photochemical/electrochemical reactivity,[11] basal plane functionalization,[10-13] and intercalant-induced chemical doping,[14] have recently appeared. Purposeful graphene modification and systematic processing, however, require a deeper understanding of graphene chemistry than is presently available.

Single atomic layer graphene is a unique two-dimensional electronic material. With all atoms on the surface, its properties are strongly influenced by the supporting substrate[15-19] and the local molecular environment.[20] While graphene shows atomic-level flatness on mica[21] and h-BN[22] substrates, it shows different degrees of local structure and corrugation when deposited on $SiO_2$/Si substrates, or when suspended across a trench.[23, 24] These structural deviations from planarity are believed to strongly affect electronic properties[25] and chemical reactivity.[26]

Adsorption on $SiO_2$/Si substrates is of central importance for technology, and such adsorption significantly modifies charge transport properties and Raman spectra.[18, 19, 27] Several independent studies have reported a significant stiffening (shift to higher energy) of the Raman G and 2D bands in thermally annealed, supported graphene;[9, 28, 29] however, the basic cause of this shift remains uncertain. In order to control and purposefully modify graphene for application, these issues need to be clearly understood.

Molecular $O_2$ exhibits a rich variety of chemical interactions with aromatic molecules[30, 31] and carbon nanotubes,[32] and adsorbed $O_2$ is a well-known, effective hole-dopant for these species.[30, 32-34] In general, distortion of aromatic π systems from planarity creates a stronger interaction with $O_2$.[31] The present study reveals that there is a subtle interplay between different types of graphene distortion and hole doping by adsorbed $O_2$. We directly correlate atomically resolved scanning tunneling microscopy (STM) images with Raman scattering measurements of hole doping. Thermal annealing induces pronounced short range distortion along with both *reversible* and irreversible $O_2$ binding and hole doping of graphene supported on $SiO_2$/Si substrates. This does not occur for naturally rippled, suspended graphene in ambient, or for exfoliated graphene in a simple loose contact with $SiO_2$/Si substrates. These observations provide a basis for understanding the wide variety of behavior previously reported for supported graphene on $SiO_2$/Si.

**Results**

**STM imaging of changes induced in graphene morphology by annealing on a $SiO_2$/Si substrate.**

Ambient STM images taken at room temperature were used to investigate annealing-induced changes in graphene morphology. To avoid contamination from resist residues, we used a shadow mask method, instead of the conventional electron beam lithography technique,[35-37] to pattern electrodes on graphene. (See Supporting Information). Atomically resolved images were observed on both the initial, directly deposited (exfoliated) samples, and on samples subjected to thermal annealing. (See the inset of Fig. 1).

Fig. 1a and 1b show large scale STM images of single-layered graphene before and after annealing at 300 °C under an Ar gas flow. Upon annealing, many dome-shaped features with a lateral size of several nm are apparent. As can be seen in the inset of Fig.1, no hint of extraneous contamination was found from extensive STM imaging on annealed graphene surface. Line profiles in Fig. 1c and 1d show that annealing increases height variations from about 0.6 nm on the initial surface to about 1.0 nm for the annealed surface, over a 150 nm range. The initial and annealed height distributions in Fig. 2a are both Gaussian with standard deviations (σ) of 0.14 nm and 0.22 nm, respectively.

Lateral autocorrelation functions[38] measured before and after annealing (see Fig. 2b) are both Gaussian near their origins, and can be parameterized with a lateral correlation length λ.[39, 40] (See Supporting Information). λ values for the initial and annealed surfaces are 2.8 nm and 5.5 nm, respectively, indicating that height correlations persist over a shorter distance in the annealed sample. Both autocorrelation functions show oscillations at low autocorrelation function G(r) values for longer distances, with the annealed surface displaying much larger amplitude oscillations than the initial surface. This oscillation (the pseudo-periodicity[40]) reflects the average separation of the surface features. The ~50% increase in $\sigma$ and the simultaneous ~50% decrease in $\lambda$ indicate that graphene undergoes significant structural deformation, leading to pseudo-periodic surface features with much higher aspect ratios after annealing. An essentially identical change was found for several spots randomly chosen within the graphene area (32 μm x 32 μm). For a double-layered sample, pronounced rippling was also observed after thermal annealing. (See Supporting Information).

**Raman spectroscopy of graphene annealed under a buffer gas atmosphere**

Raman spectroscopy is a sensitive, non-contact diagnostic for graphene crystalline quality and electrical doping. The Raman G band upshift from 1580 cm$^{-1}$ is widely interpreted as a measure of carrier doping. Actually the G band can upshift due either to the application of compressive strain,[41, 42] or to electrical doping.[43, 44] Using environment-controlled in-situ Raman spectroscopy measurements, we now show that the observed upshift in both initially exfoliated and annealed graphene is principally due to $O_2$ induced hole-doping rather than in-plane compressive strain.

The G band frequency, $\omega_G$, of a $SiO_2$/Si supported single-layered graphene sample mounted in an air-tight optical gas cell was monitored both before and after in-situ annealing in an Ar atmosphere. (See Supporting Information). Typically, $\omega_G$ is observed to be close to 1583±1 cm$^{-1}$ either under air or under Ar for an initial exfoliated graphene sample (Fig. 3*i*). This $\omega_G$ value is very close to that for intrinsic undoped graphene, indicating that air does not appreciably hole dope the initial exfoliated graphene, as previously reported.[9] Following two-hours of annealing at 290 °C under flowing Ar, $\omega_G$ measured under Ar atmosphere at 23 °C (Fig. 3*ii*) was ~1.5 cm$^{-1}$ higher than in the initial state (Fig. 3*i*). This small upshift was essentially within the range of $\omega_G$ variations measured across the sample from one spot to the next. This observed upshift, $\Delta\omega_G$, is an order of magnitude smaller than some values reported previously.[9, 28, 29] Although graphene undergoes significant structural deformation upon annealing as shown above, $\omega_G$ varies only slightly. This $\Delta\omega_G$ value sets a 0.12% upper limit for the compressive strain upon annealing, as estimated from the experimental shift rates of $\omega_G$ under uniaxial[41, 42] or biaxial[45] tensile stress. Alternatively, the slight upshift may be attributed to electron doping caused by charge transfer at the graphene-$SiO_2$ interface.[46]

Since compressive strain and direct charge transfer from $SiO_2$ are minor, the influence on $\omega_G$ of dry $O_2$ adsorption after annealing was measured consecutively at one spot (Fig. 3*iii*) as a function of $O_2$ exposure. Upon replacing flowing Ar with flowing $O_2$, $\omega_G$ upshifts by ~4 cm$^{-1}$. We conclude that $O_2$ binds on or near the annealed graphene causing a hole doping of ~2x10$^{12}$/cm$^2$.[44] This $O_2$ doping upshift could be almost completely reversed by re-exposing the graphene to flowing Ar gas (Fig. 3*iv*). The binding is reversible on a time scale of minutes. Exposure to $O_2$ (solid squares in Fig. 3*v*) following an additional two-hour annealing at 320 °C in Ar induced an even larger blueshift in $\omega_G$ (~7 cm$^{-1}$).

This observed change in $\omega_G$, however, is less than that observed previously in ambient air for annealed samples.[9] This suggests that a second species contributes to hole doping. We exposed samples annealed under Ar to subsequent wet nitrogen gas flow (Fig. 3*vii)*, instead of $O_2$. Wet nitrogen had no effect on $\omega_G$. To then test the combined effect of water and $O_2$, wet $O_2$ gas was introduced to replace the wet $N_2$ flow. Each of the measured $\omega_G$ values (diamonds in Fig. 3*viii*) was obtained from a separate spot by integrating for a period of 5 min. The combined flow induces a larger blueshift (~10 cm$^{-1}$), a value close to that observed previously in air.[9] In addition, the rate of change in $\omega_G$ is much larger than that found under pure $O_2$ (Fig. 3*v*).

Unlike the case of treatment with pure $O_2$, subsequent exposure of "wet" graphene to an Ar atmosphere (Fig. 3*ix*) leads to an incomplete recovery of $\omega_G$, a behavior that was also observed in the annealed graphene following exposure to ambient air.[9] The effects caused by water adsorption suggest that the oxygen anion produced by hole doping of graphene is more stable in the presence of water. Storage of doped samples for periods of weeks under ambient conditions also decreased reversibility in subsequent Ar flows. Note also that the G mode line-width changes significantly in response to the various gas environments, in a fashion consistent with a charge doping phenomenon.[44, 47] (See Supporting Information). In addition the annealing-induced hole doping of graphene by $O_2$ was also confirmed through electrical measurements. (See Supporting Information). All the evidence supports a conclusion that the upshift of the G band in thermally annealed graphene is caused by hole doping from $O_2$ with an assist from water molecules.

**Freestanding versus supported graphene**

To assess the effect of $SiO_2$ substrates on the annealing-induced hole doping,[9] freestanding graphene suspended in ambient across a micro-trench in the $SiO_2$ was studied. (See Supporting Information). The freestanding area of a double layer graphene sheet in Fig. 4a can be distinguished from the supported area in the optical image. The $\omega_G$ of the initial exfoliated sample was the same for the freestanding region and the supported region. Following annealing at 300 °C in an Ar/$O_2$ atmosphere, however, the supported region showed a large upshift of 6±1 cm$^{-1}$ while $\omega_G$ for the freestanding region was essentially unchanged.

This result indicates that the $SiO_2$ substrate plays an essential role in the $O_2$-induced hole doping observed upon annealing. Oxygen doping does not occur for either initial or annealed freestanding graphene, which is consistent with substrate effects in carbon nanotubes[48] and graphene.[19]

**Heavily hole-doped, un-annealed exfoliated graphene, and ultraviolet photoexcitation**

Our initial graphene samples as described above are typically nearly intrinsic with negligible doping. However, significant hole doping of un-annealed exfoliated samples has been reported[19, 49, 50] and occurs occasionally in our samples. The doping mechanism in this case was investigated as follows: In such samples, the sensitivity of $\omega_G$ to gas flow was monitored. For example, an initial $\omega_G$ value of ~1590 cm$^{-1}$ for one sample in air (Fig. 5*i*) indicates that the graphene is heavily hole-doped ($\boldsymbol{n} \sim 4 \times 10^{12}$/cm$^2$).[44] Upon replacing air with an Ar flow (Fig. 5*ii*), $\omega_G$ decreased only slightly (~1 cm$^{-1}$). This irreversible behavior is in contrast to the more completely reversible behavior of annealed samples exposed for short periods to dry $O_2$ and shown in Figure 3.

To possibly detach or neutralize dopant species, un-annealed samples were irradiated with far UV light from an Hg Pen Lamp for 10 minutes in flowing Ar. $\omega_G$ decreased substantially to ~1581.5 cm$^{-1}$ (Fig. 5*iii*), which is close to the intrinsic value observed for freestanding graphene.[19] The disorder-induced Raman D band was not detected before or after UV irradiation implying a negligible change in the sp$^2$ hybridization. Remarkably, subsequent exposure to $O_2$ led to a blueshift in $\omega_G$ (Fig. 5*iv*) which corresponds to ~70% of the initial level of doping. When the $O_2$ flow was subsequently replaced by Ar in Figure 5, $\omega_G$ decreased over time (*v*) as in the case of annealed graphene. Water by itself in wet Ar did not increase the hole density (*vi*), which is consistent with the negligible effect of water on the annealed graphene. While water transfers very little charge to graphene, it does enhance the level of doping due to $O_2$ (Fig. 5*vii*), as seen above for annealed graphene (Fig. 3). The G band blueshifts more and faster in wet $O_2$ (Fig. 5*vii*) than in dry $O_2$ (Fig. 5*iv*).

The width of the G band ($\Gamma_G$) also exhibits interesting behavior. When $\omega_G$ increases (decreases), $\Gamma_G$ decreases (increases). Such a correlation is nicely explained by a model of charge doping in graphene.[44, 47] In particular, since $\Gamma_G$ reaches the intrinsic value of charge-neutral graphene following UV irradiation,[19] the change can be attributed to un-doping by UV light. One electrical transport measurement reported a similar UV-induced un-doping but without any further investigation.[20] Note also that hole doping in annealed graphene stored for periods of weeks in ambient can be significantly removed by similar UV irradiation.

Since the response of these un-annealed graphene samples to various gases and far UV light resembles that of annealed graphene, we conclude that dopants in both cases work through the same mechanism involving $O_2$. Presumably, only a small fraction of initial exfoliated samples are conformally distorted in close contact with $SiO_2$ due to strong mechanical forces during graphene transfer. Most initial graphenes are in loose contact, as discussed below, and close contact only develops during thermal annealing. Close contact enables reversible $O_2$ binding and hole doping.

**Discussion**

**Structural distortion in graphene supported on $SiO_2$/Si substrates.**

In this work, we show that graphene distortion is minor in un-annealed exfoliated graphene resting upon $SiO_2$/Si substrates. Surface conformal distortion becomes substantial upon thermal annealing. The STM images show that thermal annealing induces dome-shaped features on graphene. These "hill"-like features have an average diameter of 2.8 nm with a 1 nm height variation. In addition, shorter lateral range distortion appears, as shown by the intense, quickly decaying peak in the autocorrelation function of the annealed sample near the origin in Fig. 2b. In general graphene sheets are believed to conform to the morphologies of underlying substrates to different degrees.[21, 36] The degree of this conformation increases under vacuum to remove ambient species trapped under or bound on graphene.[36]

Both "hills" and short λ distortion have been seen to varying extent in previous STM studies of annealed graphene on $SiO_2$/Si substrates. The "hill" features have been interpreted as graphene resting on (conformal with) high spots of the underlying $SiO_2$.[37, 51] Thermal annealing likely brings closer contact between an undulating oxide surface and the graphene sheets by driving out extraneous molecules initially trapped under the graphene. Alternatively, differential thermal expansion effects upon annealing may additionally contribute to the observed distortion.[24] In effect, graphene on $SiO_2$/Si is not a single species with well defined properties. Rather a wide range of adsorbed graphenes can be prepared, from close substrate coupling to quite loose coupling, depending upon sample processing. Close substrate coupling enables $O_2$ to hole dope graphene under ambient conditions; indeed, a second period of annealing induces greater sensitivity to $O_2$ as seen in Fig. 3.

The van der Waals interaction energy of graphene closely coupled to $SiO_2$ is significant (~6 meV/$A^2$),[36] and comparable to the elastic energy stored in graphene laterally stretched by a few percent.[52] Graphene is likely pinned at high spots of $SiO_2$ substrates by van der Waals interactions. Dome-shaped features may represent underlying oxide on hills or may be created by strain-releasing bending and buckling of graphene during thermal annealing cycles. When the sample is cooling down, the $SiO_2$ substrates (positive thermal expansion coefficient) will apply compressive strains on graphene (negative thermal expansion coefficient[24]). To release the strain, graphene buckles up and down, as shown for the case of freestanding graphene suspended across micro-sized trenches.[24] In valley areas between domes, graphene can be driven into bi-stable out-of-plane configurations by STM tips.[45, 51] All this emphasizes again that graphene on Si/$SiO_2$ can occur in a wide range of distorted configurations.

Out of plane distorted, undoped graphene will show a disorganized spatial pattern of local nm size "pools" of electrons and holes; essentially the Fermi level is not constant across the surface.[53] Vertical distortions of a fraction of a nm are calculated to produce local carrier densities of a fraction of $10^{12}$ cm$^{-2}$. These Fermi level shifts will have a direct effect on graphene electron transfer processes.

The Raman, STM, and $O_2$ chemical measurements reveal very different aspects of the behavior of graphene closely coupled to $SiO_2$/Si. The short lateral range distortion of these graphene samples implies a more significant change in the local graphene C-C bonding than occurs in the long wavelength ripples in freestanding graphene.[23, 24, 45] This distortion will weaken the π bonds and increase chemical reactivity.[54] However, this distortion (created by close coupling as seen in the STM) does *not* activate the Raman disorder D band. Binding of $O_2$ to graphene closely coupled to $SiO_2$/Si also does not activate the disorder D band. That is, neither close coupling to $SiO_2$ nor subsequent $O_2$ binding significantly distorts aromatic C atoms from $sp^2$ towards $sp^3$ hybridization.

The G band of undoped graphene is experimentally unaffected by local distortion, or by the formation of local electron and hole pools that occur upon annealing. The G band is, however, quite sensitive to the average, overall doping that occurs upon exposure to oxygen. In contrast, the 2D Raman band near 2700 cm$^{-1}$ shows a significant shape change, and the 2D/G intensity ratio decreases by a factor of 4 for graphene closely coupling to $SiO_2$/Si.[19]

**Oxygen interaction with Graphene:**

In this work both reversible and irreversible graphene doping by atmospheric oxygen have been observed. Annealed graphene is reversibly doped by brief exposure to dry $O_2$; the oxygen involved in doping is in equilibrium with gaseous oxygen in the flow cell on a time scale of minutes. Exposure to water, or storage for weeks in an ambient atmosphere, creates more irreversibly bound oxygen and doping. As shown in Fig. 5, the high doping of a minority of initially un-annealed, exfoliated samples is essentially irreversible. Irreversibly bound oxygen, not in effective equilibrium with atmospheric oxygen, can be removed by UV irradiation. After UV irradiation, subsequent brief $O_2$ exposure creates reversibly bound oxygen. Clearly more than one type of oxygen species is involved in modifying the properties of these graphene samples.

In endoperoxide structures, $O_2$ forms a modest covalent bond to a strained aromatic molecule, essentially adding across one benzene-like ring. This interaction has been extensively documented.[31] The nonplanar strain is released as $sp^2$ hybridized carbons become $sp^3$ carbons in the endoperoxide. The bonding can be reversible. There is fractional electron transfer from the aromatic species to the oxygen, thus we would expect that bound endoperoxides would dope graphene. Formation of endoperoxides is enhanced in polar solvents since polar transition states are involved. In the present work we observe that graphene loosely coupled to the $SiO_2$/Si surface (before annealing) is not doped and thus does not spontaneously form endoperoxides with atmospheric $O_2$. It may be that endoperoxide formation occurs in annealed graphene which shows increased distortion. Normally, only excited singlet oxygen forms endoperoxides; there is only one report of ground electronic state triplet $O_2$ forming an endoperoxide (with the highly strained aromatic molecule helianthrene).[55]

As discussed above the disorder D Raman band is not activated when annealed graphene undergoes either reversible or irreversible oxygen doping. The absence of the D band is in contrast to the (thermally reversible) H atom binding to the graphene basal plane, for example, which does activate the disorder D band.[10] H atom bonding produces $sp^3$ C-H bonds with a surrounding region of disturbed aromatic structure. The D band is also activated by phenyl radical bonding to the graphene basal plane; this reaction is initiated by electron transfer from graphene.[56] The absence of the D band is also in contrast to the increase in D/G intensity ratio observed in curved single-walled carbon nanotubes upon exposure to oxygen, where it appears that an endoperoxide forms.[34, 57]

In the present experiments, how intense would we expect an endoperoxide-induced D band to be? D band intensity calibrations have been recently performed in the related case of basal plane defects created by ion bombardment.[58] The D band is detectible at about $10^{11}$ defects/$cm^2$. In our present work, we observe hole densities in Fig. 5*i* of $4 \times 10^{12}$/$cm^2$. If this were attributed to endoperoxide formation, then the endoperoxide density might be ca. $10^{13}$/$cm^2$, in view of the fact that only partial charge transfer is expected. This endoperoxide density is far above the threshold for D band observation, and therefore we discount endoperoxides as the source of the observed hole doping. Nevertheless, endoperoxides do show UV photodetachment as observed in the present experiments for the more irreversible graphene doping.

The absence of the Raman D band indicates that the interaction between graphene and $O_2$ is weak. A weak, non-bonded adsorption on graphene might be occurring, or perhaps $O_2$ is bound to the $SiO_2$/Si surface near graphene after annealing. Weak, non-bonded oxygen charge-transfer complexes (donor-acceptor complexes) are well documented for π-conjugated polymers[33] and crystals.[30] These weak interactions are reversible in the cases of polythiophene[33] and pentacene.[30]

Pristine graphene has a work function about 4.5 eV, and the electron affinity of neutral gaseous $O_2$ is only about 0.4 eV; therefore electron transfer is endothermic even when the resulting Coulomb interaction is included. Thus fractional charge transfer should be minor in the ground electronic state of a non-bonded graphene-oxygen charge-transfer complex. However, Van Driel and coworkers have shown that $O_2$ binds to a clean polar silicon dioxide surface with a strong electrostatic interaction on the order of 100 meV.[59]

They demonstrated that such bound $O_2$ can accept an electron to form the anion $O_2^-$. Water would further stabilize the anion. The electron affinity of $O_2$ bound to moist silicon dioxide could be more than 0.4 eV, but still less than the graphene work function. If such a bound $O_2$ on the oxide surface were in contact with graphene, we would expect increased fractional charge transfer as compared with the direct graphene –oxygen complex described above.

In a related experiment, FET devices of carbon nanotubes on silicon dioxide exhibit electrical hysteresis in the presence of moist air, but not in dry air.[60] This hysteresis was assigned to trapped charge stabilized by water bound to the polar silicon dioxide surface, on the basis of thermal annealing and gas flow experiments. Water bound to a clean silicon dioxide surface is known to only desorb under prolonged annealing.[60]

We tentatively assign our observed graphene hole doping to partial charge transfer with $O_2$ principally bound to the silicon dioxide surface, underneath graphene. Ambient graphene hole doping has been well explained by intrinsic graphene screening of charge exchange at the graphene/$SiO_2$ substrate interface.[61] The effect of annealing is to clean the silicon dioxide surface, and to allow close approach of graphene to the surface. Irreversibility is created as water stabilizes the anionic state. It may be that bound $O_2$ diffuses to more inaccessible locations under the graphene sheet over time, adding to the observed irreversibility. Perhaps this initial species evolves into some other chemical species as well. Oxygen is critical in creating strong hole doping; calculations show that graphene directly adsorbed on clean silicon dioxide should be slightly negatively doped by electron transfer from the oxide.[46] Alternatively, the observed hole doping may be attributed to an electrochemical reaction responsible for the hole doping seen on the surface of hydrogenated diamond in the presence of ambient $O_2$ and water: $O_2 + 4H^+ + 4e^- \leftrightarrow 2H_2O$, where the electrons are provided by diamond.[62] Graphene[63] and carbon nanotubes,[64] however, are expected to be better electron donors than diamond considering their energetics with respect to electronic potential of $O_2/H_2O$ redox couple.[64] Water[65] and oxygen are expected to bind more readily on graphene lattice corrugated by annealing due to increased curvature.[66]

Fractional charge transfer requires that the $O_2$ be close enough to exchange electrons with graphene. Electron exchange rates fall off dramatically with distance. In cases where nearby $O_2$ does not effectively exchange electrons with graphene, it may be possible to create a photo-stationary state in which optically excited graphene transfers a "hot" electron to create a long lived $O_2^-$ anion. In this way anions might be formed from physisorbed $O_2$ on the outer graphene surface. Consistent with this possibility, in pentacene transistors hole doping induced by oxygen increases upon visible light exposure. It may be that the Raman laser itself is partially creating a photo-stationary state with $O_2$ molecules not directly in contact with graphene.

**Spontaneous hole doping in exfoliated graphene**

Our work clearly shows that the Raman G band upshift in annealed graphene is caused by hole doping instead of compressive strain,[29] and the hole doping is mainly due to the $O_2$ molecule, but not to dopants from the dry and degassed $SiO_2$ substrates.[46]

As C. Casiraghi et al. have documented,[49] pristine exfoliated graphene sheets prepared on $SiO_2$ substrates under ambient conditions have wide distributions of G band energy and linewidth due to hole doping. While electrical measurements[43, 50] have confirmed this spontaneous hole doping, its cause has remained unknown. Our study shows that heavily hole-doped pristine graphene can be almost completely un-doped by irradiating with UV light under an Ar atmosphere and that subsequent exposure to $O_2$ gas recovers the initial doping level. Similarly, graphene hole-doped by annealing can be un-doped by UV light.[9] Based on the similar responses to UV light, $O_2$ and water vapor, we conclude that the heavy hole doping in unannealed graphene has the same origin as that of annealed graphene; namely formation of oxygen-graphene complexes where oxygen moieties withdraw electrons from graphene. The broad distribution in the doping level of as-prepared pristine graphene samples can be attributed to the widely varying extent of structural deformation; because of varying roughness of $SiO_2$ substrates and the random nature of the mechanical exfoliation/deposition process, the extent of corrugation can be expected to vary from sample to sample.[35, 36]

**Conclusion**

We have shown that that the Raman G band upshift in annealed graphene on silicon dioxide is caused by hole doping and not compressive strain. There are two independent factors controlling the doping: (1) the degree of graphene coupling to the substrate, and (2) exposure to oxygen and moisture. By direct comparison of STM images of graphene taken before and after annealing cycles, we show that thermal annealing induces a pronounced distortion in graphene supported on $SiO_2$. Graphene Raman spectra show both reversible and irreversible oxygen doping of this structurally deformed graphene closely coupled to silicon dioxide at room temperature. Water vapor does not dope graphene noticeably, yet it greatly promotes hole doping caused by $O_2$. We tentatively assign the electron acceptor to $O_2$ electrostatically bound to the clean silicon oxide surface.

**Acknowledgements:** This work was supported at Columbia University by the National Science Foundation through grant CHE-07-01483 (to G.W.F.), by the National Science Foundation through the NSEC Program (CHE-06-41523), by the Department of Energy (DE-FG0298ER-14861 to L.E.B.) and by the New York State Office of Science, Technology, and Academic Research (NYSTAR). Material support was provided by the U.S. Department of Energy (DE-FG02-88-ER13937 to G.W.F.) and by the Air Force Office of Scientific Research (MURI FA955009-1-0705). This research was also supported by Basic Science Research Program through the National Research Foundation of Korea (NRF) funded by the Ministry of Education, Science and Technology (2009-0089030) (to S.R.) and International Cooperation of Science and Technology (Global Research Laboratory program) (to Y.Y. and P.K.). We thank Michael Steigerwald, Andrew Crowther and Naeyoung Jung for insightful comments.

**Supporting Information Available**

Description of experimental methods, thermally induced morphology change of double layer graphene, definition of correlation length (λ), electrical confirmation of annealing-induced hole doping. This material is available free of charge via the Internet at http://pubs.acs.org.

**Figures and Captions**

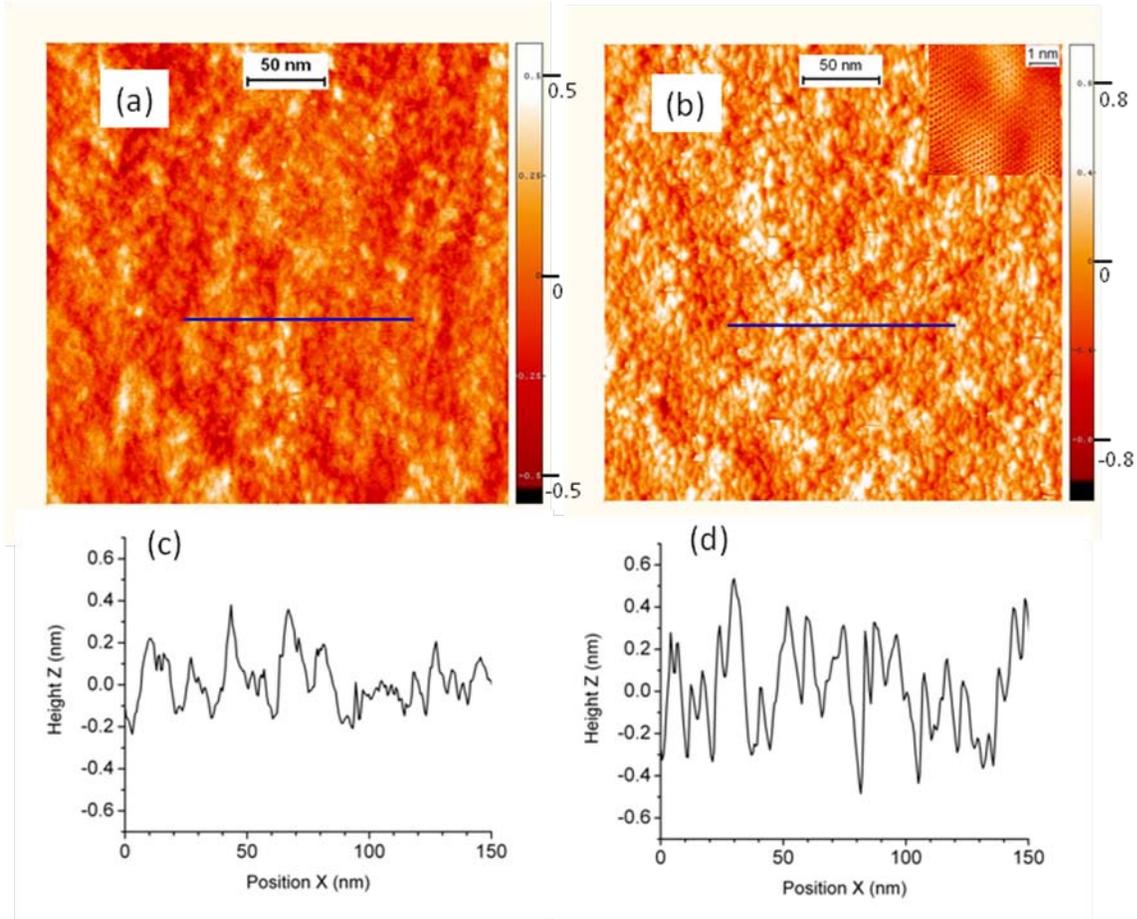

**Fig. 1**. STM images of thermally induced ripples in single layer (1L) graphene. (a) A typical large-scale STM image of an initial 1L graphene sample. (b) A typical large-scale STM image of the 1L sample after thermal treatment. The initial 1L sample was annealed in an Ar gas flow at 300 °C for 2 hr. Inset: a zoom-in STM image showing the honeycomb structure of 1L graphene. (c) and (d) Line profiles along the blue lines shown in (a) and (b) respectively. All STM measurements were carried out under ambient conditions in the constant current mode ($V_{bias}$ = 0.5 V, $I_{tunnel}$ = 0.5 nA).

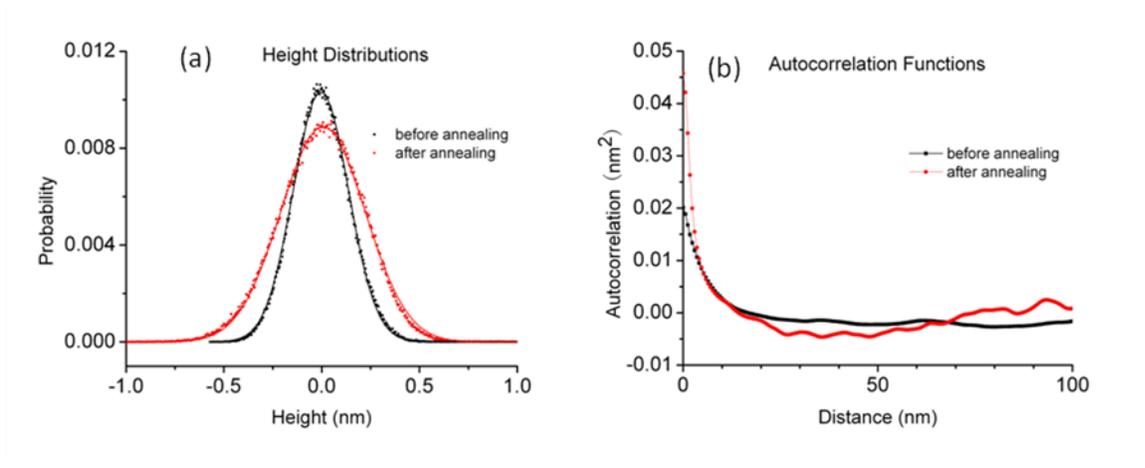

**Fig. 2**. Statistical analysis of the annealing-induced graphene morphology change. (a) Height distributions of initial (black dots) and annealed (red dots) graphene, acquired from the whole areas of Fig.1a and 1b. The distributions can be fitted by Gaussian functions with standard deviations of 0.14 nm and 0.22 nm for initial (black line) and annealed graphene (red line), respectively. (b) Autocorrelation functions, G(r), acquired from the whole areas of Fig.1a (initial graphene) and Fig. 1b (annealed graphene). Near the origin both autocorrelation functions can be fitted by Gaussian functions with autocorrelation lengths ($\lambda$) of 2.8 nm and 5.5 nm for initial (black dots) and annealed (red dots) graphene. The oscillatory pattern found for G(r) of the annealed graphene reflects the pseudo periodicity of the annealing-induced ripples. See the text for details.

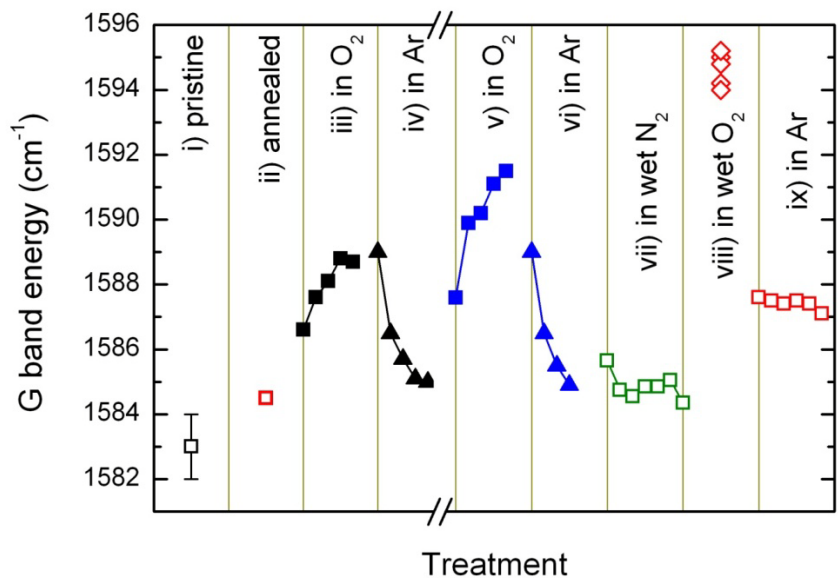

**Fig. 3**. Oxygen sensitivity of 1L graphene rippled by annealing and monitored by Raman G band energy. Raman measurements were taken at room temperature: i) in Ar atmosphere before any thermal treatment, ii) in Ar following a two-hour annealing in Ar at 290 °C, iii) in $O_2$, iv) in Ar, v) in $O_2$ following an additional two-hour annealing in Ar at 320 °C, vi) in Ar, vii) in $N_2$ containing water vapor, viii) in $O_2$ containing water vapor, and ix) in Ar. Each measurement was performed at a new spot located at least 1.5 μm away from previous spots in the graphene sample except the series of points: each series of the points in iii) through vii), and ix) corresponds to consecutive measurements at one spot. Each of the five dots in viii) was also obtained from a separate spot on the graphene. See the Supporting Information for the corresponding variations in the G band linewidth.

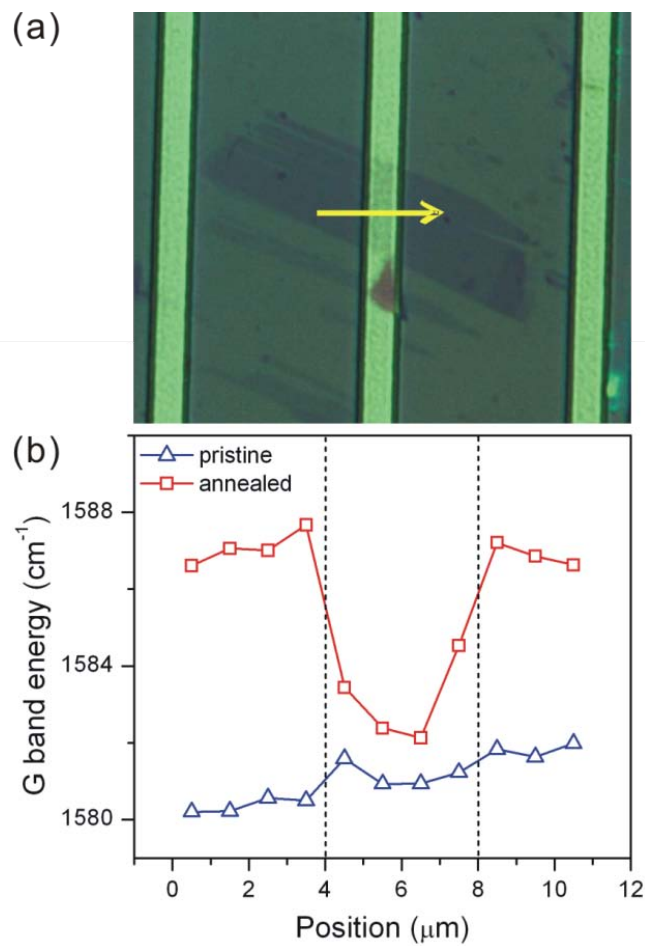

**Fig. 4**. (a) Optical micrograph (45 μm x 38 μm) of 2L graphene suspended across a 4 μm-wide trench (bright vertical stripes). Freestanding graphene suspended over the central trench exhibits less contrast than the supported area. (b) Raman G band energy ($\omega_G$) measured along the yellow arrow across the central trench. Dotted vertical lines correspond to the edges of the trench. The Raman measurements were done under ambient conditions before and after annealing in Ar:$O_2$ at 300 °C for two hours.

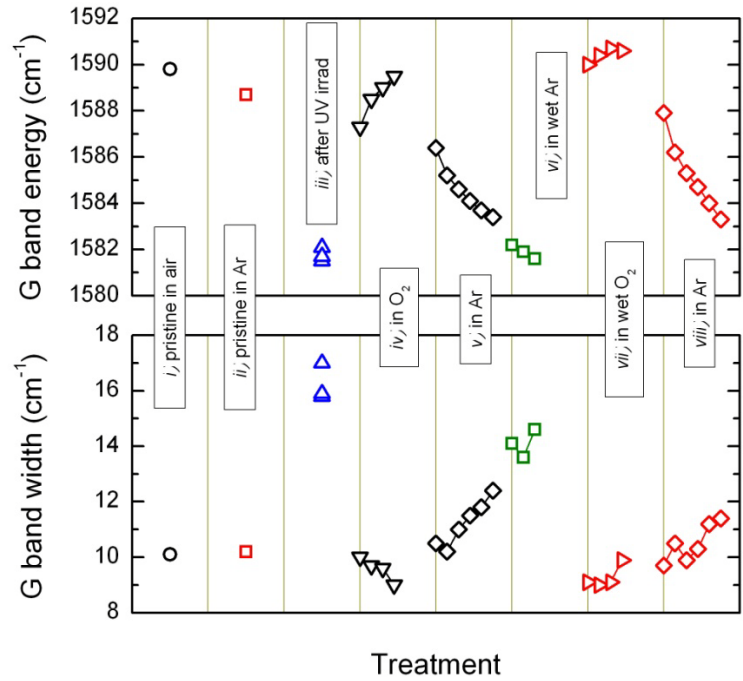

**Fig. 5**. Oxygen sensitivity of a pristine 1L graphene sample with a high level of spontaneous hole doping ($n \sim 4 \times 10^{12}/cm^2$), monitored by Raman G band energy and linewidth. Raman measurements were taken at room temperature: **i)** in air before any treatment, **ii)** in Ar before any treatment, **iii)** in Ar following a 10-min irradiation with UV light from a Hg lamp in Ar, **iv)** in $O_2$, **v)** in Ar, **vi)** in Ar containing water vapor, **vii)** in $O_2$ containing water vapor, **viii)** in Ar. Each of the three dots in **iii)** was obtained from a separate spot in the graphene. Each series of the points in **iv)** through **viii)** corresponds to consecutive measurements at one spot. The line width includes an instrumental broadening of 3.5 cm$^{-1}$.

**TOC Figure**

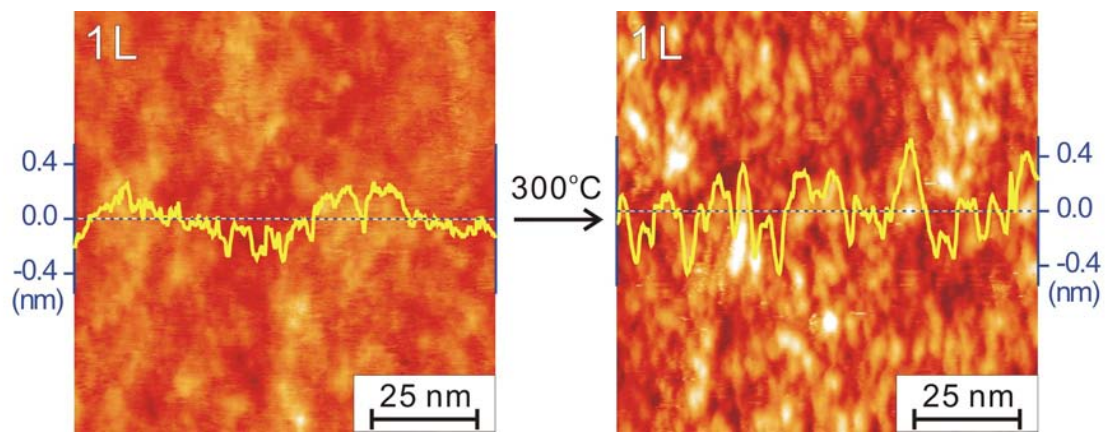